# A realist view of the canonical EPRB experiment based on quantum theory and its consequences

**Boon Leong Lan**

School of Engineering & Science

Monash University

P.O. Box 8975

46780 Kelana Jaya, Selangor

Malaysia

## Abstract

A realist view of the Einstein-Podolsky-Rosen-Bohm experiment with spins based on quantum theory is presented. This view implies that there is no action at a distance. It also implies that the measurement result *A* (*B*) for particle *1* (*2*) depends on *both* magnet angles, and hence the probability of obtaining the result *A* (*B*) also depends on *both* magnet angles. In light of these realist implications, it is clear that what is wrong at least with local realistic theory is not the locality or no action-at-a-distance assumption itself but rather the formal implementation of that assumption.

The recent Einstein-Podolsky-Rosen-Bohm (EPRB) experiments of Weihs *et al* [1] with photon polarizations, which closed the locality loophole, and Rowe *et al* [2] with ion energies, which closed the efficiency loophole, both confirmed the prediction of quantum theory instead of the prediction of local realistic theory *á la* Bell. Although both loopholes are yet to be closed in a single experiment, it is expected [1,3] that such a definitive experiment will also agree with quantum theory. If so, the experimental violation of the prediction of local realistic theory would imply that at least one of the assumptions of the theory (in addition to locality or 'no action-at-a-distance' [4] and realism, there are other assumptions [5] as well) is inconsistent with nature. In this paper, I will present a realist view of the canonical EPRB experiment with spins based on quantum theory. The implications of this realist view, as we will see, allow us to pinpoint what is wrong with local realistic theory.

In the canonical EPRB experiment (see figure 1), a source produces a system of two spin-half particles (labeled *1* and *2*) that fly apart in opposite directions, each towards a Stern-Gerlach magnet. Each magnet can be rotated in a plane perpendicular to the line of flight of the particles: $\theta$ ($\phi$) gives the direction of magnet *1* (*2*). Let $|\pm,\theta\rangle$ ($|\pm,\phi\rangle$) be the eigenstates of the projection of the spin operator of particle *1* (*2*) onto the unit vector in the direction of magnet *1* (*2*). The spin part of the quantum wave function for the system, known as the singlet state in the literature, can be expressed [6] in terms of the set of product states $\{|\pm,\theta\rangle|\pm,\phi\rangle\}$ that I call system spin states:

$$|S\rangle = \frac{1}{\sqrt{2}}\left[-i\sin\left(\frac{\theta-\phi}{2}\right)|+,\theta\rangle|+,\phi\rangle \;+\; \cos\left(\frac{\theta-\phi}{2}\right)|+,\theta\rangle|-,\phi\rangle\right.$$

$$-\cos\left(\frac{\theta-\phi}{2}\right)|-,\theta\rangle|+,\phi\rangle \;+\; i\sin\left(\frac{\theta-\phi}{2}\right)|-,\theta\rangle|-,\phi\rangle\Bigg] \qquad (1)$$

The quantum conditional probabilities for the possible joint measurement outcomes are easily determined from the expansion coefficients in equation (1):

$$P(++|\theta,\phi) = P(--|\theta,\phi) = \frac{1}{2}\sin^2\left(\frac{\theta-\phi}{2}\right) \qquad (2)$$

$$P(+-|\theta,\phi) = P(-+|\theta,\phi) = \frac{1}{2}\cos^2\left(\frac{\theta-\phi}{2}\right), \qquad (3)$$

where, for instance, $P(+-|\theta,\phi)$ is the probability of measuring a spin up at magnet *1* and a spin down at magnet *2* given that magnet *1* set at angle $\theta$ and magnet *2* is set at angle $\phi$.

Prior to measurement in an experimental run, because the magnet angles could be chosen after the two particles have been created, one of the four possible system spin states for each possible pair of magnet angles must have existed in the system since the creation of the two particles. For a chosen pair of magnet angles, measurement reveals the pre-existing system spin-state for that pair of angles. For different experimental runs, the pre-existing system spin-state for a given possible pair of magnet angles will generally be different, the four possible states occur in accordance with the quantum probabilities given by equations (2) and (3).

To illustrate the realist view presented above, table 1 gives a partial list of the pre-existing system spin states for a few pairs of possible magnet angles, one system state per pair of angles, in a hypothetical experimental run. If the magnets are set at angles $\theta'$ and $\phi'$ respectively, then measurement will yield a spin up at magnet *1* and a spin down at magnet *2*, revealing that that the pre-existing system spin-state for this pair of angles is $|+,\theta'\rangle|-,\phi'\rangle$. However, if the setting of the magnet angles are $\theta''$ and $\phi''$ respectively, then the pre-existing state $|-,\theta''\rangle|-,\phi''\rangle$ will be revealed by measurement. And so on.

The realist view I have presented implies that, in an experimental run, there isn't any action at a distance whatsoever because measurement merely reveals the pre-existing system spin-state for the chosen pair of magnet angles. Furthermore, because the pre-existing system spin-state varies with *both* magnet angles, the measurement result $A$ for particle *1* also depends on *both* magnet angles and the measurement result $B$ for particle *2* also depends on *both* magnet angles:

$$A(\theta,\phi) \text{ and } B(\theta,\phi). \tag{4}$$

The dependence of $A$ and $B$ on the magnet angles above respectively implies that the probability $P_1$ of obtaining the result $A$ also depends on *both* magnet angles and the probability $P_2$ of obtaining the result $B$ also depends on *both* magnet angles:

$$P_1(A|\theta,\phi) \text{ and } P_2(B|\theta,\phi). \tag{5}$$

In light of the realist implications above, it is clear that locality or no action-at-a-distance (I) *does not* require that $A$ does not depend on $\phi$ and $B$ does not depend on $\theta$:

$$A(\theta) \text{ and } B(\phi), \tag{6}$$

and it (II) *does not* require that $P_1$ does not depend on $\phi$ and $P_2$ does not depend on $\theta$:

$$P_1(A|\theta) \text{ and } P_2(B|\phi), \tag{7}$$

contrary to what Bell [7] and others [8] had assumed. (II) was also previously recognized by Jaynes [9] and Kracklauer [10], based on the consideration of the interpretation of conditional probabilities. Thus, what is wrong at least with local realistic theory is not the locality assumption itself but rather the formal implementation of that assumption in equation (6) or equation (7) by Bell and others.

For other EPRB experiments, a realist view that is based on the quantum-mechanical description can also be constructed in each case, leading in all cases to the same conclusions as in the spin case. In particular, there is no action at a distance, contrary to widespread popular belief, see for example [11-17], that experiments which have been performed prove the existence of an instantaneous action at a distance. Jaynes [9] and Kracklauer [10], for instance, have also maintained that there is no action at a distance.

**Acknowledgement**

My thanks to A. F. Kracklauer for bringing his paper and Jaynes' paper to my attention at the Garda 2001 workshop.

**Captions**

**Table 1**

A partial list of the pre-existing system spin-state versus possible setting of magnet angles in a hypothetical experimental run.

**Figure 1**

A schematic diagram of the canonical EPRB experiment. A source, located at the origin, produces two spin-half particles, labeled *1* and *2*, which fly towards two oppositely located magnets set at angles $\theta$ and $\phi$ respectively.

**Table 1**

| Magnet Angles | Pre-existing System Spin-State |
|:---:|:---:|
| $\theta'$, $\phi'$ | $\left|+,\theta'\right\rangle\left|-,\phi'\right\rangle$ |
| $\theta''$, $\phi''$ | $\left|-,\theta''\right\rangle\left|-,\phi''\right\rangle$ |
| $\theta'''$, $\phi'''$ | $\left|+,\theta'''\right\rangle\left|-,\phi'''\right\rangle$ |

**Figure 1**

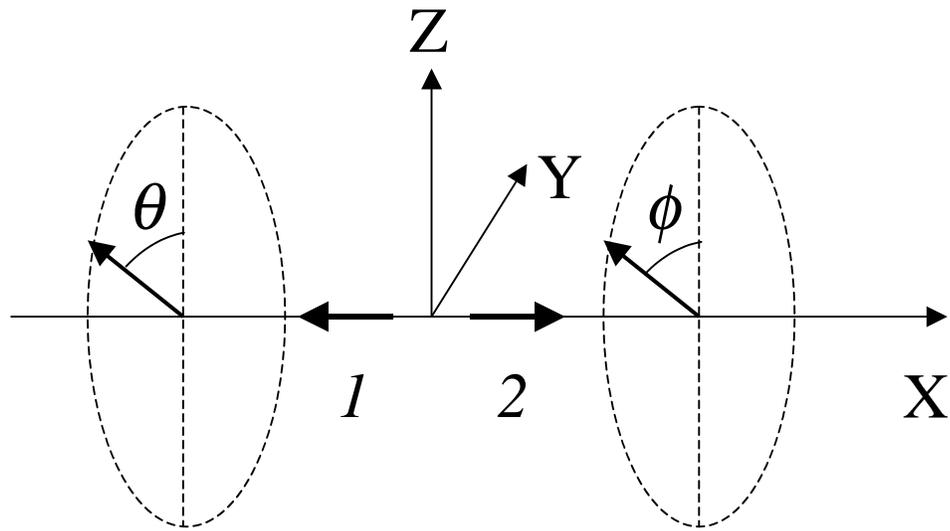